\documentclass[aps,prd,twocolumn,showpacs,superscriptaddress,groupedaddress]{revtex4-1}  
\usepackage{amsmath}
\usepackage{graphicx}
\usepackage{color}

\newcommand{\beq}{\begin{equation}}
\newcommand{\eeq}{\end{equation}}

\begin{document}

\title{Detecting Stellar Lensing of Gravitational Waves with Ground-Based Observatories}
\author{Pierre Christian$^{1}$, Salvatore Vitale$^{2}$, and Abraham Loeb$^{1}$}
\affiliation{$^{1}$Astronomy Department, Harvard University, 60 Garden St., Cambridge, MA 02138\\
$^{2}$LIGO, Massachusetts Institute of Technology, Cambridge, Massachusetts 02139, USA}
\begin{abstract}
We investigate the ability of ground based gravitational wave observatories to detect gravitational wave lensing events caused by stellar mass lenses. We show that LIGO and Virgo possess the sensitivities required to detect lenses with masses as small as $\sim 30 M_\odot$ provided that the gravitational wave is observed with a signal-to-noise ratio of $\sim30$. Third generation observatories will allow detection of gravitational wave lenses with masses of $\sim 1 M_\odot$. Finally, we discuss the possibility of lensing by multiple stars, as is the case if the gravitational radiation is passing through galactic nucleus or a dense star cluster.
\end{abstract}

\maketitle

\section{Introduction}

The recent Laser Interferometer Gravitation-Wave Observatory (LIGO) discoveries of gravitational waves from black hole binaries \citep{LIGO1,2016PhRvL.116x1103A, 2017PhRvL.118v1101A,2017PhRvL.119n1101A,2017ApJ...851L..35A} opened a new frontier for the study of astrophysical objects using gravitational radiation. Much like electromagnetic (EM) radiation in classical astrophysics, gravitational radiation can be lensed by massive objects. Lensing of gravitational radiation in linearized General Relativity can be computed with the same techniques as those employed in the familiar lensing of EM waves.    

Much of the previous literatures on gravitational lensing of gravitational waves (GWs) focused on lensing in the geometric optics limit \citep{2010PhRvL.105y1101S, 2013JCAP...10..022P}, where the wavelength of gravitational waves is small compared to the spatial scale of the lenses and ray optics is sufficient. The lenses responsible for much of the optical depth in this regime are galaxies, which split the gravitational wave signal into copies separated by a time delay of order a few months. 

The stellar lensing events that are considered in this work preempts these strong gravitational lensing by the lens host galaxy. As a beam of gravitational radiation passes through the galaxy, it will first be split into two beams due to it being strongly lensed by the galactic potential. Each of these beams can then be lensed by stars in the galaxy. Because the time delay between the arrival of the two beams is $10 - 100$ days for third generation gravitational wave observatories and $\lesssim 1$ day for LIGO \citep{2018MNRAS.480.3842O}, at first only one beam will be detected on Earth. If the first of these beams are found to also be lensed by stars, we can be confident that the beam passes through the core of a galaxy. If this is the case, then it will also be strongly lensed by the galaxy, and thus we can expect that in the close future, the strongly lensed copy of the original beam will arrive on Earth.

In the wave-optics regime, the Laser Interferometer Space Antenna (LISA) \citep{LISA} possesses the necessary sensitivities to extract the lens' mass from a lensed signal \citep{2003ApJ...595.1039T}. The goal of our study is to show that such observations can also be achieved by ground based observatories.  

The sensitivities of advanced LIGO~\citep{2015CQGra..32g4001L} and advanced Virgo~\citep{TheVirgo:2014hva} to lensing by intermediate mass black holes (IMBH) are explored in Ref \citep{LiTBD}, which found that current generation observatories are capable of detecting the lensing IMBH with $98\%$ confidence. Further, Ref \citep{LiTBD} found that LIGO and Virgo can distinguish between a point mass lens and a singular isothermal lens provided that the redshifted lens mass is $200 M_\odot$.

In this work, we focus on the capabilities of ground based GW detectors to detect stellar mass lenses. While stellar mass lenses are more numerous than IMBHs, the amplitude of their lensing signal is much smaller. This requires us to extend our study to include upcoming third generation gravitational wave detectors. 

\section{Background and notations}
We consider GWs in the perturbed Friedmann-Lema"tre-Robertson-Walker (FLRW) metric, written in terms of the conformal time $\eta$,
\beq
ds^2 = a^2 \left[ -\left( 1 + 2 U \right) d\eta^2 + \left( 1 - 2 U \right) d \mathbf{r}^2  \right] \; ,
\eeq
where $\mathbf{r}$ is the spatial coordinate, $U$ is the gravitational potential of the lenses, and $a$ encodes the universal scale factor. Considering linear perturbation on this metric,
\beq
g_{\alpha \beta} = g^B_{\alpha \beta} + h_{\mu \nu},
\eeq
where $h_{\mu \nu}$ is separated into $\phi$, its amplitude and $e_{\mu \nu}$, its polarization, $h_{\mu \nu} = \phi e_{\mu \nu}$, one can obtain that the equation of motion for $\phi$ is simply given by the wave equation,
\beq
\partial_\mu \left( \sqrt{-g^B} g^{\mu \nu}_B \partial_\nu \phi  \right)  = 0 \;,
\eeq
where $g^B$ is the determinant of the metric. In Fourier space, $\tilde{\phi}(f, \mathbf{r})$, the equation reads
\beq \label{eq:wave}
\left( \nabla^2 + \tilde{\omega}^2  \right) \tilde{\phi} = 4 \tilde{\omega}^2 U \tilde{\phi} \; ,
\eeq 
where $\tilde{\omega} = 2 \pi f$ is the GW frequency. Following \citep{2003ApJ...595.1039T}, we define the amplification factor $F(\omega, \mathbf{r} )$ as the ratio between the lensed and unlensed $\tilde{\phi}$. 

The setup of our problem consists of three parallel planes, called the \emph{source}, \emph{lens}, and \emph{observer} planes. The angular diameter distances along the normal from the observer plane to the source and lens planes are labelled $D_S$ and $D_L$, respectively, while the distance between the source and lens planes is labelled as $D_{LS}$. GWs are emitted by a point in the source plane, travel freely to the lens plane, where they are lensed by a gravitational potential $U$ that is assumed to be localized in the thin (width $\ll c/f$) lens plane, before reaching the telescope at the observer plane. 

Coordinates can be set up on the three planes. We use the notation of Ref \citep{2013IJAA....3....1N, 2003ApJ...595.1039T}, where $\mathbf{\xi}$ is the coordinate at the source plane, $\mathbf{\eta}$ is the coordinate at the lens plane, and $\mathbf{\delta}$ is the coordinate in the observer plane. We also employ the following dimensionless coordinates,
\begin{align}
\mathbf{x} &= \frac{\mathbf{\xi}}{\xi_0}  \; , \nonumber
\\ \mathbf{y} &= \frac{D_L}{D_S} \frac{\mathbf{\eta}}{\xi_0}  \; , \nonumber
\\ \mathbf{d} &= \left( 1 - \frac{D_L}{D_S} \right) \frac{\mathbf{\Delta}}{\xi_0}  \; , \nonumber 
\end{align}
where $\xi_0$ is some characteristic length-scale defined by $\xi_0 = D_L \theta_E$ where $\theta_E$ is the Einstein angle for a point mass lens,
\begin{equation}
\theta_E^2 = \frac{4 G M_L}{c^2} \frac{D_{LS}}{D_L D_S} \; ,
\end{equation}
where $M_L$ is the lens mass. Furthermore, we will work with the dimensionless frequency, which for the point mass lens is
\beq
\omega = \frac{4 G M_L (1+z)}{c^3} \tilde{\omega} \; ,
\eeq 
where $z$ is the lens redshift. From this point on we will adopt units where $G=c=1$.

Using this setup, the solution of Equation (\ref{eq:wave}) is given by the Fresnel-Kirchhoff integral,
\beq \label{eq:Fresnel}
F(\omega, \mathbf{y}) = \frac{\omega}{2 \pi i} \int d^2 x \exp{\left[  i \omega T(\mathbf{x}, \mathbf{y})  \right]  }\; ,
\eeq
where the time delay function, $T(\mathbf{x},\mathbf{y})$ is given by 
\beq
 T(\mathbf{x},\mathbf{y}) = \frac{1}{2} \left( \mathbf{x} - \mathbf{y} - \mathbf{d} \right)^2 - \Psi(\mathbf{x}) \; ,
\eeq
with $\Psi$ being the lensing potential.

\subsection{Wave optics lensing}
Integrating Equation (\ref{eq:Fresnel}) with the stationary phase method is valid when the wavelength is much smaller compared to the characteristic scale of the lens. This condition requires, 
\beq
\omega \gg 1 \; .
\eeq
A detector operating in the frequency band of LIGO, with a characteristic frequency of $f \sim 100$ Hz, is capable of detecting lenses where $\omega \ll 1$. In this regime, the geometric optics approximation breaks down and one has to integrate Equation (\ref{eq:Fresnel}) in full. 

\begin{figure}
\centering
\includegraphics[width=3.4in]{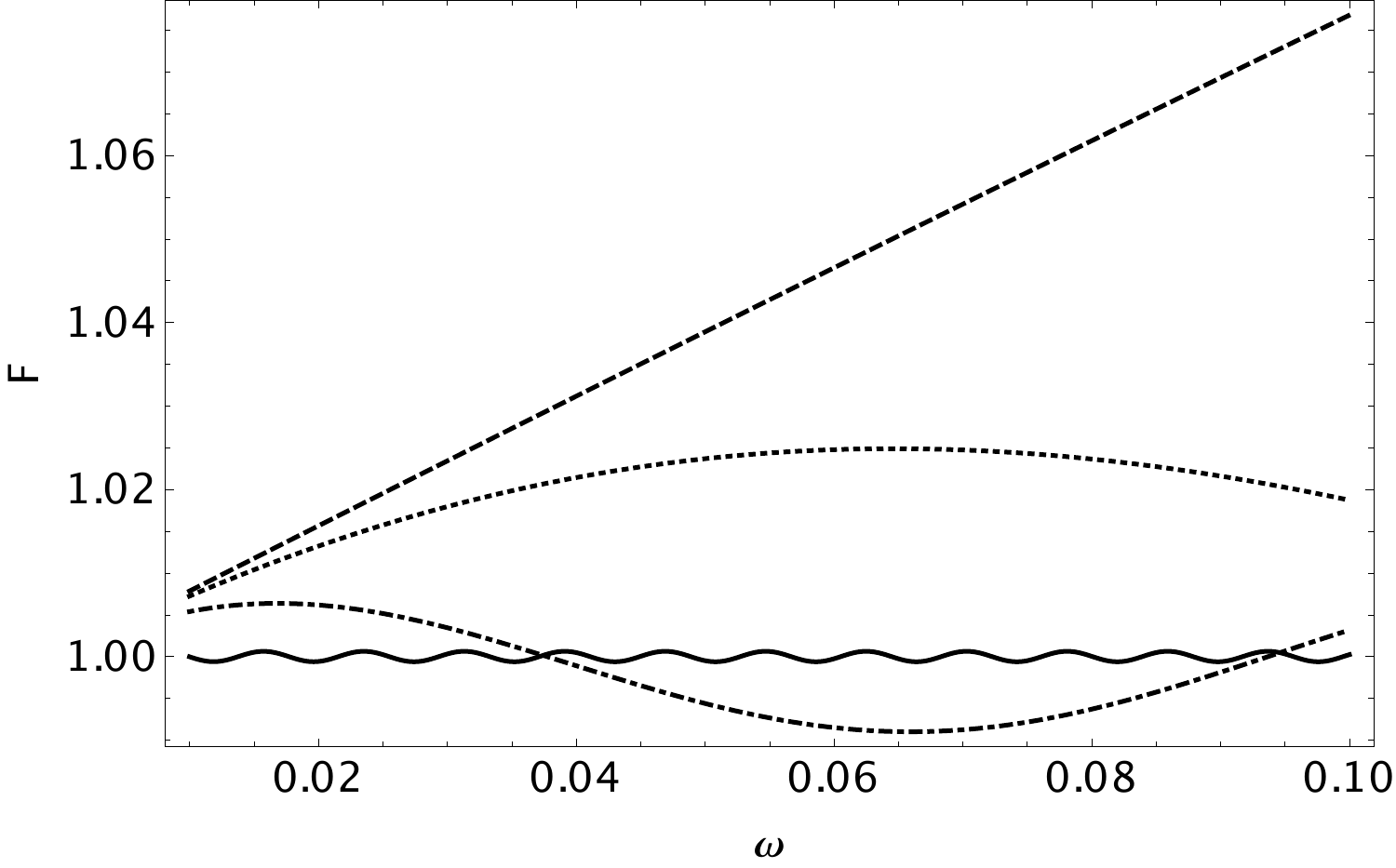}
\caption{The amplification as a function of $\omega$ for a point mass lens where $y=1, 5, 10, 40$ (dashed, dotted, dot-dashed, and solid). When the position of the source projected to the lens plane is small ($y \sim 1$), one can obtain amplification that is $\sim$linear in $\omega$. In the LIGO band ($\omega \sim 0.01-0.1$ for a solar mass lens) this results in a deviation from the unlensed signal of a few percent. As the distance increases, the amplitude of the deviation becomes smaller.}
\label{fig:WavePointMass}
\end{figure}

For a point mass lens located at $\mathbf{\eta}=0$, $\psi(\mathbf{x}) = \log | \mathbf{x} |$, and Equation (\ref{eq:Fresnel}) integrates to \citep{1974PhRvD..9.2207,2003ApJ...595.1039T},
\begin{align}
F(\omega) &= \exp{\left\{  \frac{\pi \omega}{4} + i \frac{\omega}{2} \left[ \log{ \left(  \frac{\omega}{2} \right)} - 2 \phi_m(y) \right]  \right\} } \Gamma \left(1 - \frac{i}{2} \omega \right) {} \nonumber \\
&\;\;\;\;\;\;\; \times _1F_1 \left( \frac{i}{2} \omega, 1 ; \frac{i}{2} \omega y^2 \right) \;,\label{Eq.Fomega}
\end{align}
where $y \equiv |\mathbf{y}|$ is the dimensionless impact parameter, 
\beq
\phi_m(y) = (x_m - y)^2/2 - \log x_m \; ,
\eeq
and
\beq
x_m = \frac{y + \sqrt{y^2+4}}{2}\; .
\eeq
The amplification for $\omega \sim 0.01-0.1$ and a variety of $y$ values is plotted in Figure \ref{fig:WavePointMass}. For a frequency of $100$Hz, this corresponds to lenses of mass $1-10 M_\odot$. 

\begin{figure}
\centering
\includegraphics[width=3.3in]{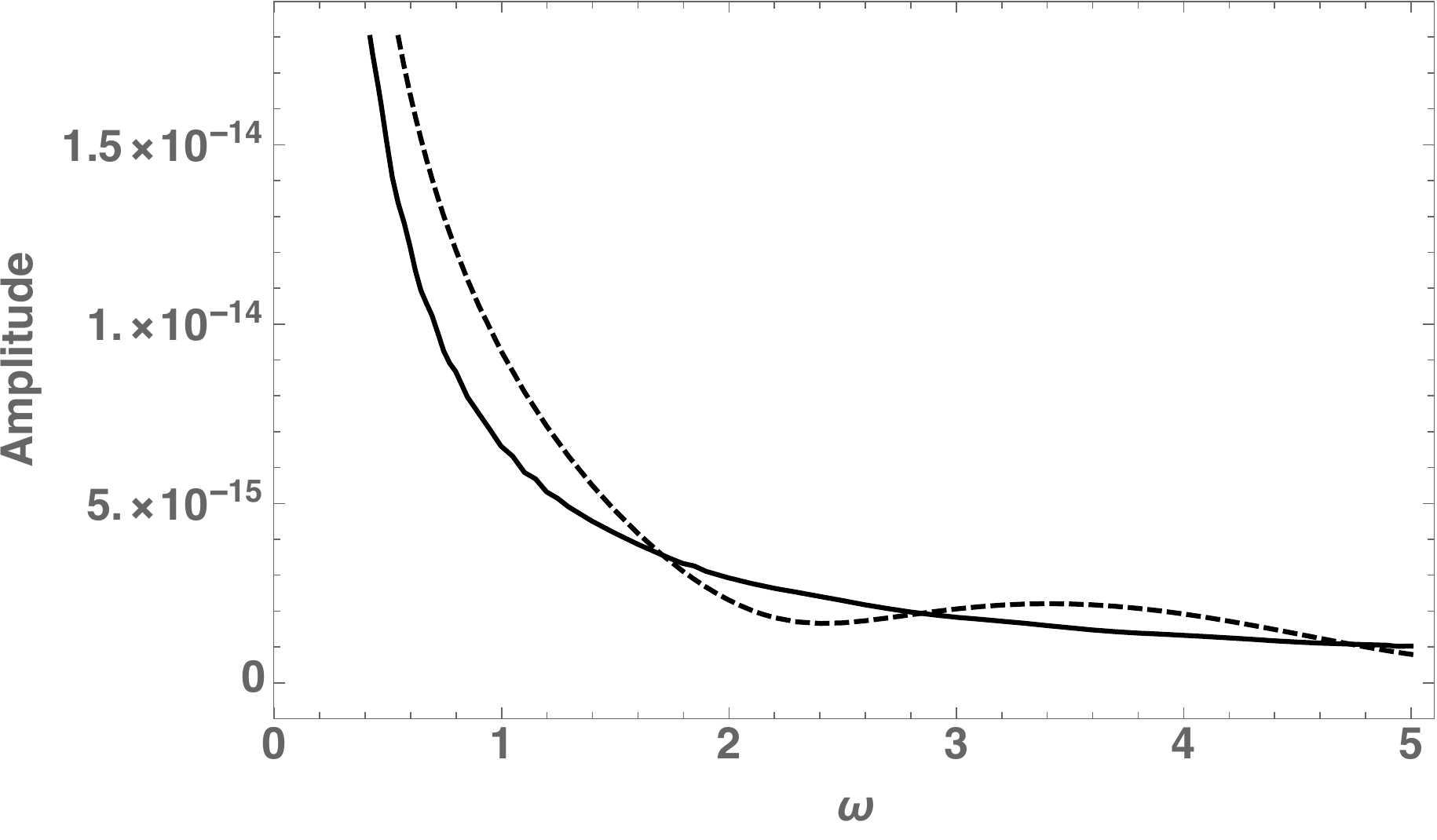}
\caption{The amplitude of waveforms from a $30+30 M_\odot$ binary at $1$ parsec lensed by a $30 M_\odot$ lens. We exemplify a strongly lensed event with a lens at $y=1$ (dashed) and a weakly lensed event with a lens at $y=10$ (solid). An unlensed signal would sit on top of the $y=10$ line. Note that in addition to the the oscillating features at high frequency that is characteristic of wave interference, there is an overall amplification of the signal.}
\label{fig:lensed_waveform}
\end{figure}

Two examples of lensed waveforms are shown in Figure \ref{fig:lensed_waveform}. The unlensed waveforms are PhenomA phenomenological models during the inspiral \citep{2007CQGra..24S.689A, 2018arXiv180301944C}. While more complex phenomenological models exist, the PhenomA model suffices for this illustrative purpose.

\section{Method}
We have modified the parameter estimation algorithm currently used by the LIGO and Virgo collaborations~\cite{2015PhRvD..91d2003V,2016PhRvL.116x1102A} to allow for the presence of a lens along the line of sight to the source.
This is a stochastic sampler that explores the parameter space and produces posterior distributions for the unknown parameters on which the gravitational-wave signal depends. In \emph{absence} of a lens, these include masses and spins of the two compact objects, the sky position, distance, orientation and polarization of the source, as well as the time and phase at coalescence~\cite{2016PhRvL.116x1102A}. Throughout this work, we use the effective-precession waveform \textsl{IMRPhenomPv2}~\cite{2014PhRvL.113o1101H}.

Our lens model allows for two extra parameters: the mass of the lens, $M_L$, and the parameter $y'\equiv y M_L$, which are sampled together along the other.
Once a waveform corresponding to the unlensed signal is generated, $h_{\mathrm{Unlensed}}(\vec{\theta})$, the two lens parameters are used to calculate Eq.~(\ref{Eq.Fomega}), which yields the lensed signal $h_{\mathrm{Lensed}}(\vec{\theta},M_\mathrm{lens},y')=F(M_\mathrm{lens},y') h_{\mathrm{Unlensed}}(\vec{\theta})$, where $\vec{\theta}$ are the Compact Binary Coalescence (CBC) parameters in absence of lens.

While the physical parameters of the lensing model are $M_L$ and $y$, we used the parameter $y'$ to smoothly handle the no-lens case without discontinuities or ill-defined parameters. The parameter $y'= y M_L$ is trivially correlated to the lens mass, $M_L$. However, as they enter different parts of the waveform models and not always in the same combination, they are not degenerate.

Given a GW signal (real or simulated) the algorithm can be run with the lens parameters (``Lens'' model) to measure or put an upper bound on the lens mass. After the evidence~\cite{Jaynes:2003jaq} for both the ``Lens'' and ``No lens'' model is calculated, one can compute the odds ratio defined as 
\beq \label{eq:odds}
\rm{Odds} = \frac{P(\rm{lens}|\rm{data})}{P(\rm{no \; lens}|\rm{data})} \; .
\eeq


\section{Results}
\subsection{Current Generation Observatories}
We ran our code on simulated signals observed by LIGO and Virgo with signal-to-noise (SNR) values of $15, 30,$ and $60$, where we injected lenses of $0, 1, 10, 20, 30, 60$ and $100 M_\odot$ with an impact parameter of an Einstein angle, $\theta_E$. The masses of the simulated CBC signal are compatible with heavy binary source similar to GW150914.

\begin{figure}
\centering
\includegraphics[width=3.3in]{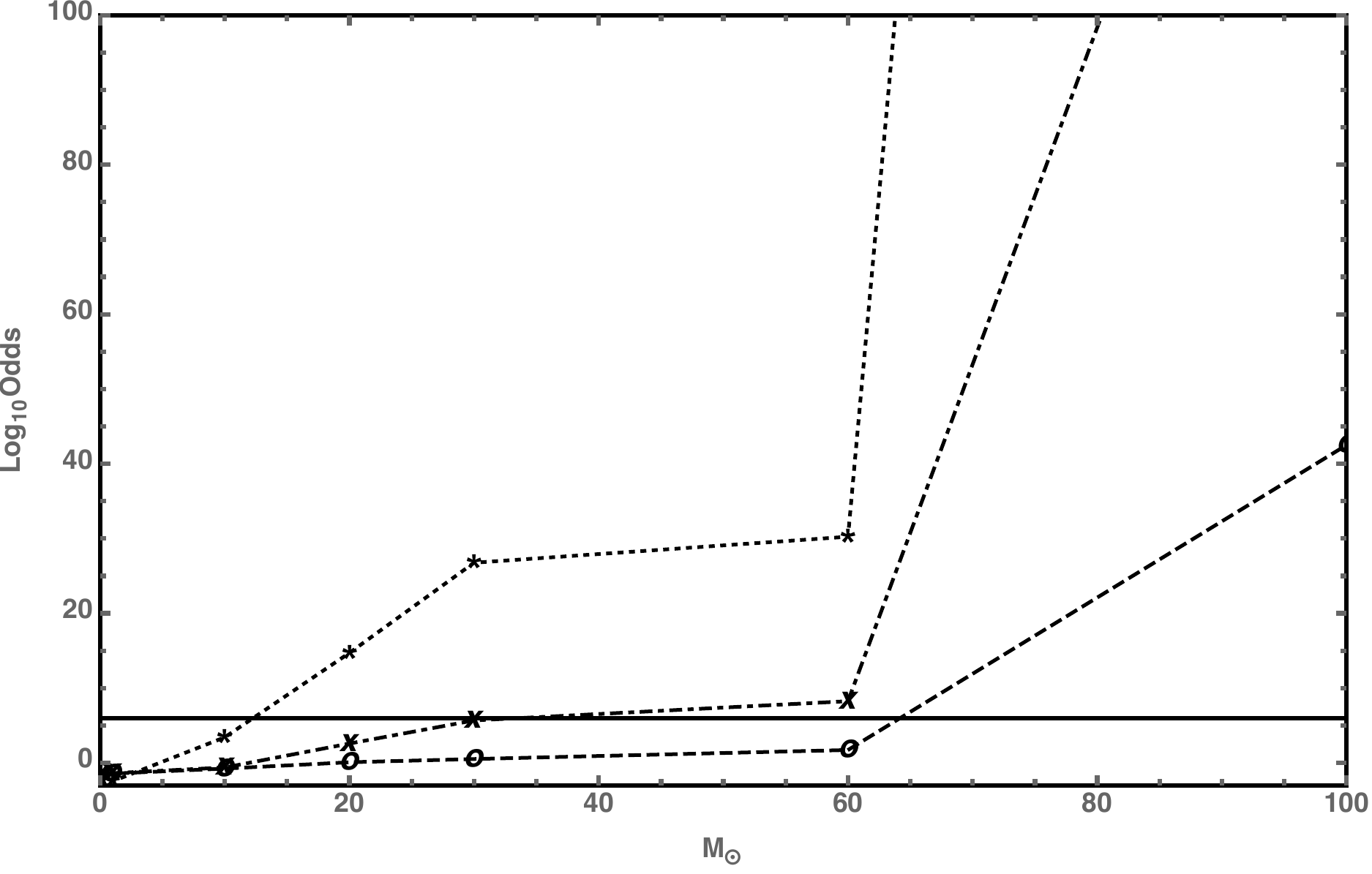}
\caption{The odds ratio computed as defined by Equation (\ref{eq:odds}). The circles, crosses, and stars denote SNR values of $10, 30,$ and $60$, respectively. Dashed, dot-dashed, and dotted lines connecting the points are drawn to guide the eye. Odds values $>1$ indicate that the lensed model is preferred over the unlensed model. The solid horizontal line indicates a 3$\sigma$ detection.}
\label{fig:CurrentGen}
\end{figure}

%
The results are plotted in Figure (\ref{fig:CurrentGen}). At a SNR ratio of $30$, which is moderately high for current generation observatories, lenses can be detected at $>3\sigma$ when they possess masses larger than $\sim 30 M_\odot$. Higher SNR events allowed smaller lenses to be detected. At SNR$=60$, lenses as small as $\sim 10 M_\odot$ can be detected. LIGO and Virgo can potentially detect smaller lenses if the impact parameter is significantly smaller than an Einstein radius, but such cases are expected to be rare.

If the mass function of black holes follows the mass function of massive stars, it is reasonable to expect the mass function of binary black holes to be bottom-heavy. Assuming that binary black holes are uniformly distributed in space, this bottom-heavy mass function means that SNR$>60$ events would be rare. However, there is the possibility that most of LIGO black holes are \emph{macro}lensed by intervening galaxies \citep{2018arXiv180205273B}. Such a macrolensing event would lower the mass requirement for a binary to be the gravitational wave source of an SNR$=60$ event, while allowing said binary to be located at a larger distance. This would enhance the rate of SNR$>60$ events and thus open the possibility of LIGO detecting stellar-mass lensing events.


As a straightforward application, we ran our algorithm on the stretch of public LIGO data containing the gravitational wave event GW150914~\citep{LIGO1,LOSC-GW150914}. We found that the waveform detected for GW150914 is consistent with a lens mass of $M=0$, i.e. GW150914 is most probably not a microlensed event with an upper bound 90\% confidence interval for the lens mass of $50 M_\odot$.

\subsection{Third Generation Observatories}
\begin{figure}
\centering
\includegraphics[width=3.2in]{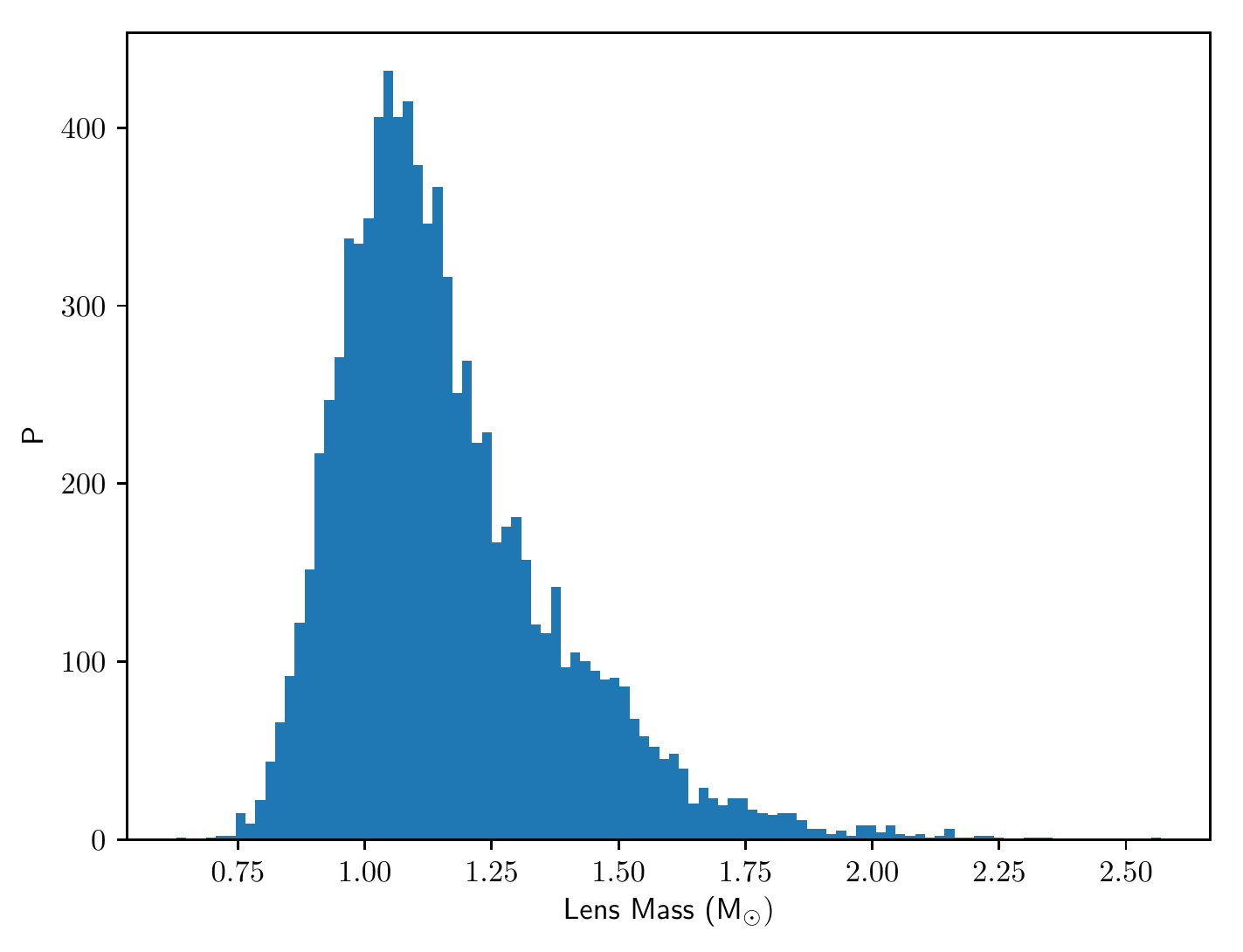}
\caption{The resulting posterior distribution of lens mass for an event with an $M=1 M_\odot$ lens observed by a third generation observatory. SNR is set to $3000$ and impact parameter is an Einstein radius. Note that a vanishing lens mass is clearly excluded.} 
\label{fig:ThirdGenMass}
\end{figure}

Proposed ground based third-generation (3G) GW observatories, such as the Einstein Telescope~\cite{2010CQGra..27s4002P}
 and the Cosmic Explorer~\cite{Evans:17}, allow detections of BBH events from high redshift~\cite{Evans:17,2017PhRvD..95f4052V,EinsteinTelescope}, and will detect nearby events with SNR of hundreds or thousands~\cite{2016PhRvD..94l1501V}. Such high SNR events can potentially allow much smaller lenses to be detected. To show this, we ran our algorithm on a simulated GW150914-like source as observed by a third generation observatory, with an injected lens of $1 M_\odot$. Figure \ref{fig:ThirdGenMass} shows the resulting posterior distribution for the lens mass. As seen in Figure \ref{fig:ThirdGenMass}, 3G observatories can detect lenses as small as $1 M_\odot$. As there are many more lenses with such masses than those with masses of $\sim 30 M_\odot$, we expect that detection of lensing events by stellar mass lenses will be mostly confined to 3G detectors. If the lenses obey the mass function for stars, the number of lensed events that we can expect from a 3G detector is greater than that of current generation detectors by a factor $q$, where
\begin{align}
q &= \frac{R_{\rm3G}}{R_{\rm CG}} \frac{\int_{1 M_\odot}^{\infty} m^{-2.3 } dm }{\int_{30 M_\odot}^{\infty} m^{-2.3 } dm  } \approx \frac{R_{\rm3G}}{R_{\rm CG}} 100 \; ,
\end{align}
where $R_{\rm3G}$ is the overall rate of gravitational wave detection by 3G observatories, $R_{\rm CG}$ is the overall rate of gravitational wave detection by current generation observatories, and we have taken the stellar mass function to be of the Salpeter form \citep{1955ApJ...121..161S}. Note that we have not fully explored the lower limit of the masses of the lenses that will be detectable by third generation observatories. It is likely that these observatories will detect lenses with masses even smaller than $1 M_\odot$.

\section{Rate estimate}

The rate of stellar lensing is intrinsically tied to the rate of strong lensing, as the surface number density of stars in the core of galaxies is large enough that if a beam passes within this core, there is an order unity chance that it will pass within the Einstein radius of a star. The number of stellar lensing events per year, $N_*$, is therefore
\beq
N_* ~\sim N_g *\left( \frac{h_c}{h_E} \right)^2 \; ,
\eeq
where $h_c$ is the angular size of the galaxy core, $h_E$ the Einstein angle associated with the galaxy potential, and $N_g$ the number of galaxy (strong) lensing events per year. 

The number of galaxy lensing events per year for a 3G telescope has been calculated to be over 100 events per year, where for most of these events the source is a binary black hole \citep{2013JCAP...10..022P, 2014JCAP...10..080B, 2015JCAP...12..006D}. Plugging in the numbers for a Milky Way Equivalent Galaxy, as well as putting the source at $z \sim 2$ and the lens halfway to the source, we obtain that $(h_c/h_E)^2 \sim 0.01$. This gives the number of stellar lensing events per year to be $N_* \sim$ few. However, this number is extremely conservative. This is because most of the lensing optical depth is provided by galaxies that are much more massive than the Milky Way. 

We can scale the factor $(h_c/h_E)^2$ as follows,
\beq
\left( \frac{h_c}{h_E} \right)^2 \sim \frac{R_c^2}{M_g} \; ,
\eeq
where $R_c$ is the radius of the galaxy core and $M_g$ the mass of the galaxy. The relation between galactic luminosity and core radius has been found to be $L \sim R_c^{0.84}$ \citep{1985ApJ...292..104L}. Therefore, 
\beq
R_c \sim L^{1.19} \sim M_g^{2.38} \; ,
\eeq
where in the last relation we have used the Faber-Jackson relation. Therefore,
\beq
\left( \frac{h_c}{h_E} \right)^2 \sim M_g^{3.76}  \; .
\eeq
Therefore, the larger the lensing galaxy, the more probable it is for the gravitational wave beam to also be stellar lensed. This is pertinent especially because as mentioned before, most lenses are massive galaxies. 

Because most sources are expected to be at $z \sim 2$, the typical detection will not have an SNR of thousands. However, because gravitational wave amplitude only goes down as the distance, the difference in SNR between a source at $z \sim 0.1$ and $z \sim 2$ is only $\sim 20$, leaving an SNR in the hundreds. This is plenty of SNR to detect small lenses with masses of a few solar masses, because as seen in Figure \ref{fig:CurrentGen}, $SNR \sim 60$ is enough to provide a $\sim 3 \sigma$ detection of $\sim 10 M_\odot$ lenses.

\section{Wave optics lensing by multiple masses}
The centers of galaxies are dense enough that the Einstein rings of stars (on an angular scale of $\sim 1 \mu$arcsecond) can overlap. Indeed, the probability for there to be another star an Einstein radius, $\xi_0$, away from a particular star is \citep{1993ASPC...50..141P, 2015ApJ...798...78C}
\beq
P \approx 1 - \exp{\left[ - \sigma \pi \xi_0^2 \right]  } \;, 
\eeq 
which approaches unity for a stellar mass density, $\sigma$, corresponding to $\sim 1\; \rm g \; cm^{-2} \sim 4.8 \times 10^9  M_{\odot}/\rm{kpc}^2$.

In this regime, it is important to understand the effects of lensing by multiple masses. Assuming that the lensing happened in a thin plane, the lensing potential by $N$ point masses is given by
\beq 
\psi(\mathbf{x}) =  \sum_{i}^N \log{\left| \mathbf{x} - \mathbf{x_i} \right|} \; .
\eeq
To simplify our calculation, we employ the fact that distant lenses do not affect the signal by imposing a cutoff on $\psi(\mathbf{x})$. In particular, we will ignore any lenses that are more than an Einstein radius away from the source,
\beq \label{eq:superposelens}
\psi(\mathbf{x}) \approx \sum_{i}^N H(\left| \mathbf{x}-\mathbf{x_i} \right|) \log{\left| \mathbf{x} - \mathbf{x_i} \right|} \; ,
\eeq
where $H(\left| \mathbf{x}-\mathbf{x_i} \right|)$ is a tophat kernel that is unity when $\left| \mathbf{x}-\mathbf{x_i} \right|$ is less than an Einstein radius, and zero otherwise. In doing so, we do not need to include all the point masses in the lensing galaxy in $\psi(\mathbf{r})$, but only the lenses whose Einstein rings intersect. Obviously, this number depends on the surface density number of stars in the lensing galaxy. 


\begin{figure}
\centering
\includegraphics[width=3.3in]{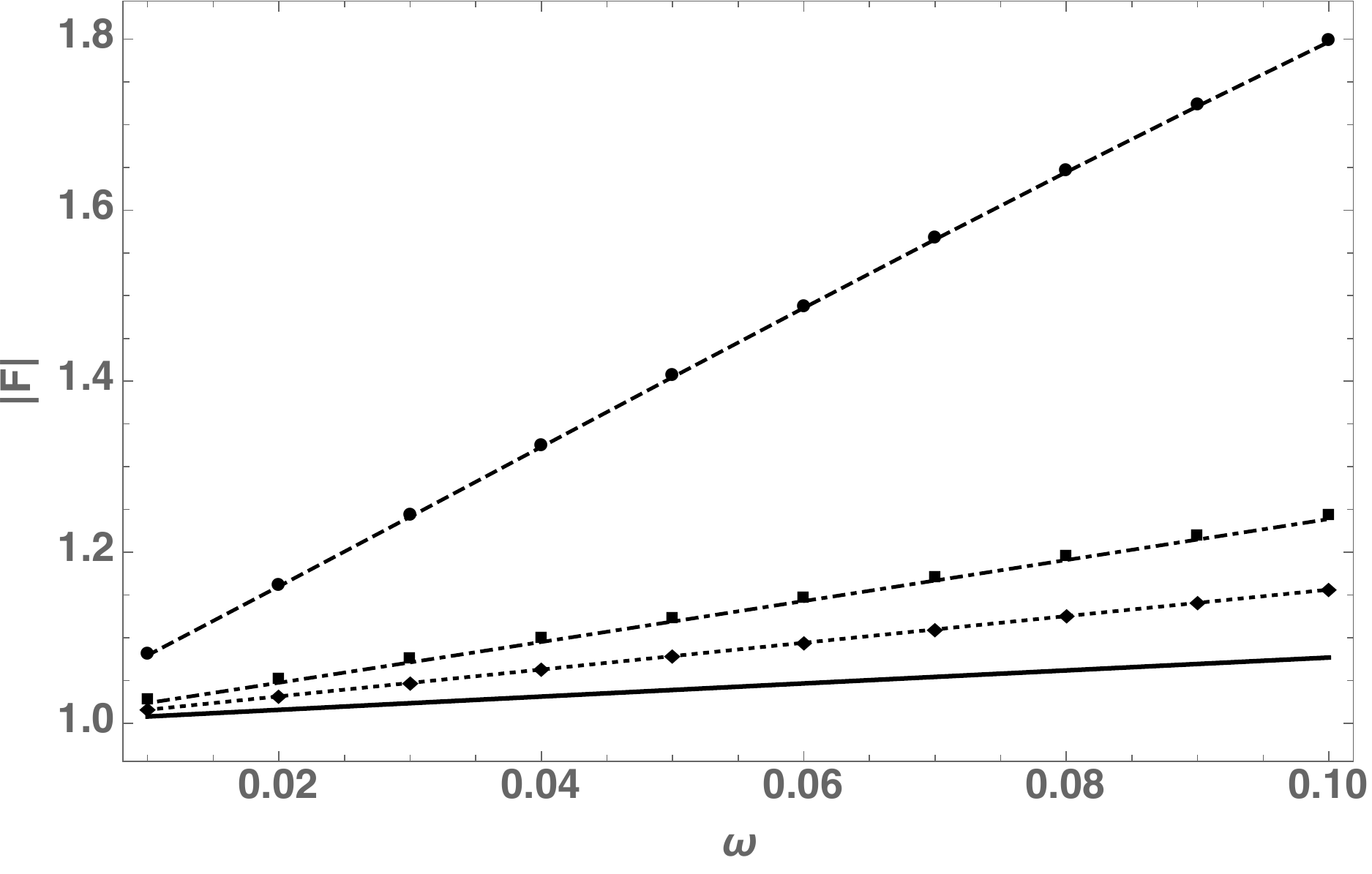}
\caption{The amplification factor due to lensing as a function of $\omega$ for $1$ (analytical solid line), $2$, $3$, and $10$ (dotted, dot-dashed, and dashed, respectively) point mass lenses. For $1 M_\odot$ lenses, the $\omega$ range corresponds to the LIGO frequency range. The lenses are distributed randomly, but consistently in the lens plane, so that the two lens case corresponds to the single lens case plus a randomly distributed second lens, and similarly for the $3$ and $10$ lenses cases. The position of the source in the source plane is $(0, 1)$, and the position of the observer in the observer plane is $(0,0)$ in Einstein angle units. In this regime where $F(\omega)\propto \omega$, more lenses generally generate a larger lensing effect.}
\label{fig:MultLens}
\end{figure}

The upper limit on the stellar surface mass density in a dense system is $\Sigma_{\rm max} \sim 10^{11} M_{\odot}/\rm{kpc}^2$ \citep{2010MNRAS.401L..19H}. Assuming that most of the stellar mass is in stars of mass $\sim 1 M_{\odot}$, this gives a surface number density of $\sim 10^5 \; \rm{pc}^{-2}$. Using the fact that at cosmological distances, the Einstein angle of such stars are $\sim 1 \mu$arcsecond, and that stars are randomly distributed in the lensing plane, we created realizations of star fields in the lensing plane. Even for such a dense system, the number of Einstein ring intersections is of order a few. The expected number of overlapping Einstein rings, $N_O$, can be estimated as follows. If $A_1=1/\sigma$ is the area where only one star is expected, then 
\begin{align} 
N_O &= \frac{\pi (2 \xi_0)^2}{A_1} =  \pi (2 \xi_0)^2 \sigma  
\\ &= 4.3 \times \left( \frac{\sigma}{10^5/\rm{pc}^2} \right) \; .
\end{align}
We therefore expect GWs to be significantly lensed by only a few lenses. However, as a Poisson process, $N_O$ is Poisson distributed. Therefore, for a $\sigma=10^5 \; \rm{pc}^{-2}$ system, there is a $\sim1$\% chance for the beam to interact with $\sim10$ lenses. 

To this end we calculate the lensing amplitude $F(\omega)$ by numerically integrating Equation (\ref{eq:Fresnel}) using a Levin method integrator \citep{2008mgm..conf..807M}. The resulting magnification amplitudes for $2$, $3$, and $10$ lenses are plotted in Figure (\ref{fig:MultLens}). For $1 M_\odot$ lenses in the LIGO band, $F(\omega)$ scales linearly with $\omega$. The general trend is that more lenses yield a larger deviation in amplitude. This means that if a lensing event is detected at low $\omega$, it might be difficult to distinguish between lensing by a single point mass or lensing by multiple point masses. However, if a larger range of $\omega$ is observed, these two models will differ significantly.

The computation in this section assumes that the lensing potentials of multiple stars can be superposed in the manner of Equation (\ref{eq:superposelens}). In reality, nonlinear effects can produce caustics in the source plane. Further, these caustics can overlap and create large magnifications on scales larger than the Einstein radius. This effect has recently been observed for a star lensed by a galaxy cluster \citep{2018NatAs...2..334K} which allowed constraints on compact dark matter to be placed \citep{2018ApJ...857...25D}. We leave the computation of caustic curves in the wave optics regime to a future work.

\section{Conclusions}

We have shown that in order for current generation GW observatories to detect gravitational wave lensing events, a lens mass of at least $\sim 30 M_\odot$ is required, provided that the gravitational waveform is detected at a signal to noise of $\sim 30$. If the gravitational wave source is weaker or is located further away, this number will increase correspondingly. We also note the possibility that the gravitational wave is macrolensed by an intervening galaxy in addition to the stellar lensing event. This could potentially produce an event with large enough SNR for LIGO to detect stellar lenses.

Furthermore, we have shown that 3G detectors can detect lenses of masses as small as $1 M_\odot$. Since $1 M_\odot$ lenses are much more numerous than $\sim 30 M_\odot$ lenses, many more lensing events will be detected by third generation detectors than current generation detectors.

\begin{acknowledgements}
This work was supported in part by the Black Hole Initiative at Harvard University, which is funded by a grant from the John Templeton Foundation.
\end{acknowledgements}

\bibliography{BibFile.bib}

\end{document}